# A Bayesian semi-parametric model for small area estimation

Donald Malec[1] and Peter Müller[2]

*U.S. Census Bureau and M. D. Anderson Cancer Center*

**Abstract:** In public health management there is a need to produce subnational estimates of health outcomes. Often, however, funds are not available to collect samples large enough to produce traditional survey sample estimates for each subnational area. Although parametric hierarchical methods have been successfully used to derive estimates from small samples, there is a concern that the geographic diversity of the U.S. population may be oversimplified in these models. In this paper, a semi-parametric model is used to describe the geographic variability component of the model. Specifically, we assume Dirichlet process mixtures of normals for county-specific random effects. Results are compared to a parametric model based on the base measure of the Dirichlet process, using binary health outcomes related to mammogram usage.

## Contents

## 1. Introduction

Large national surveys are generally constructed to provide estimates for a diverse group of data users. Estimates, from a single survey, often cover scores of topics and are usually provided for as many population groups as the sample size will support. Population groups of interest consist of both large and demographic groups and subnational areas. As the data users are thought to hold a variety of prior opinions, estimates for a particular group are usually constructed using only sample data from the group in question along with randomization theory. Often estimates are needed for small population subgroups that do not contain enough data for precise estimation. In this situation, small area estimation methods, based on statistical models and parametric exchangeability assumptions that use data from similar groups are often used. These exchangeability assumptions are generally applied across small areas if there is no a priori evidence to the contrary. Of course, this assumption may be incorrect and, without alternatives to this assumption, will remain throughout the inference.

---

[1]Statistical Research Division, U.S. Bureau of the Census Washington, DC, USA, e-mail: donald.j.malec@census.gov

[2]Department of Biostatistics, M. D. Anderson Cancer Center, Houston TX 77030, USA, e-mail: pmueller@mdanderson.org

We propose a semi-parametric, hierarchical model for small area estimation. The model produces estimates for a small area that selectively use data from only subsets of other small areas. The uncertainty about which subsets are valid is automatically accounted for. This model is useful when, *a priori*, one suspects that all small areas are not equally similar but one does not know where the dissimilarities are. The model is an extension of currently used parametric hierarchical models for small area estimation. Although the parametric hierarchical models are flexible enough to adjust for the amount of "borrowing," it is required that the subsets where "borrowing" can take place be prespecified. In practice, subset identity may not be known. In addition, some small areas may be outliers and should not be used in drawing inference about other areas and, vice versa, other areas should not be used in drawing inference about them. The semiparametric model, proposed here, will adapt to such outliers. The desired flexibility is accomplished by using a Dirichlet process (DP) prior to define a probability model on possible partitions of small areas. Specifically, the DP prior defines partitions of the set of all counties. Conditional on an assumed partition the model assumes a homogeneous population within all counties in a partitioning subset. This alleviates the difficult problem of specifying a large number of prior probabilities for possible partitions as in Malec and Sedransk [9]. In addition, the DP model is computationally tractable. The model is applied to the National Health Interview Survey (NHIS) for estimates of mammography utilization.

We begin with a historical overview of methodological improvements for small area estimation in Section 2. This is followed, in Section 3, by a description of the NHIS and how the variables were selected for the model of mammography utilization. In Section 4 we motivate and describe the prior model on the random effects. Section 5 discusses details of the implementation by Markov chain Monte Carlo simulation. In Section 6 we apply the model to the NHIS and compare results with alternative approaches.

## 2. Early and current methods for small area estimation

Early estimates for small areas were based on the assumption that an outcome is primarily a function of demographic class. By knowing the demographic characteristics of a small area and estimating the national-level prevalence of outcomes within the same demographic groups, one can derive small area estimates. Estimates based on this assumption are known as "synthetic estimates." One of the first publications using this method pertained to synthetic state estimation of disability [14].

A systematic, empirical study of biases in synthetic estimation was undertaken by Schaible et al. [20]. Little data is usually available to evaluate the bias of actual small area estimates. However, using a wide range of outcomes available in the U.S. decennial census, it was shown that the basic homogeneity assumptions underlying synthetic estimation were frequently unwarranted. To alleviate this problem, the biases in synthetic small area estimates can be modeled as random effects. Related estimates of this type have been proposed in Fay and Herriot [3], Dempster and Tomberlin [2] and Battese and Fuller [1]. The model-based approach using random effects opens up the possibility for a variety of models, resulting in realistic estimators that may combine data in a nonlinear manner.

Much of the recent development in small area estimation has concentrated on obtaining estimates from random effect models without resorting to asymptotic approximations using Bayesian Inference and Markov chain Monte Carlo methods (see, e.g., [10]). In current research, the distributional assumptions of the random effects models are being questioned. For example, alternative models which allow for spatial effects have been employed by Ghosh et al. [5] and others. Maiti [7] uses finite mixture models for random effects.

For a recent more detailed review of small area estimation, see Rao [16].

## 3. The NHIS and the selection of variables for estimation

The NHIS is typical of many national, representative, personal interview surveys. A personal interview can be relatively costly but response rates are generally higher than for other modes of interview and complicated questions, especially questions involving visual cues, can be asked. A substantial contributor to the total cost of a personal interview survey is travel cost. To minimize travel costs, each interviewer usually conducts interviews near his or her residence. Hence, sampling units are usually small geographic land areas. As a result, there is a relatively large sample size in these selected areas but relatively few areas selected across the country. In addition, the overall sample size is selected to produce traditional, randomization-based survey estimates at the national level. Due to high sampling costs, the sample size is usually inadequate to make precise design-based estimates for most small areas. However, estimates are usually desired at smaller levels, for example, for allocation of funds or for assessing and administering health services. The required precision of these estimates is usually of the same order of magnitude as that required for national estimates, resulting in an extreme shortfall of data for making small area estimates.

The NHIS is redesigned every ten years following the decennial census, taking advantage of up-to-date population data. The NHIS is based on a complex sample design of clusters of housing units, called segments. Segments are selected from within a Primary Sampling Unit (PSU), which is usually a group of contiguous counties. PSUs are sampled from strata, constructed to contain PSUs with similar socio-demographic content. See Massey et al. [11] for details. The segment is the smallest sampling unit for many questions asked in the NHIS. Subsamples, within segments, are also collected on special health topics from either a subsample of households or from people within households. The questions on mammography usage, studied here, are from a subsample of housing units, within segments sampled during years 1993 and 1994.

The sampled outcomes from the NHIS are not an i.i.d. sample from the population. They should not be considered a simple random sample due to both non-response and to possible selection bias due to the sampling design. Nonresponse is accounted for by poststratifying by county, age and race groups. Sample selection effects are countered by including design variables in the model. An example of an alternative approach, in which the sample selection process is modeled directly, is given by Malec et al. [8].

In this paper we consider inference related to one of the questions in the NHIS which asks whether a mammogram had been obtained within the past two years. Periodic mammograms for screening of breast cancer are highly recommended for women of certain ages. Estimation of the mammography utilization rate by age and race for local areas is important for evaluation purposes. Although NHIS data based on single years of age by race within each county is available, as are census population estimates for counties by race and five-year age groups, we chose to evaluate the $D = 6$ demographic groups defined by the three age groups (30–39, 40–49, 50+) crossed with race (black, non-black).

We consider a model:

$$P(y_{idk} = 1) = p_{id}, \tag{1}$$

where $y_{idk} = 1$ if the $k$-th person in demographic group $d$ and county $i$ did receive a mammogram in the past two years, and $y_{idk} = 0$ otherwise. It is assumed throughout that there may be differences between local effects of mammography utilization due to demographic groups. This is why six different prevalence rates, $p_{id}$, $d = 1, \ldots, D$, are used for each county $i$. The first step in the model selection process is to select county covariates to account for small area effects. The county variables used are from the Area Resource File [22]. Twenty-two county covariates were evaluated. These covariates comprised regional status of the county, urban variables, percent of population working in white collar jobs, percent of population working in agriculture and construction, urban status, unemployment rate, percent of households that are renters, percent minority (black, asian, hispanic and mexican), degree of education, population density, and median home value. Let $x_i$ denote a vector of all possible county-level covariates and let $\mu_d$ denote demographic effects, $d = 1, \ldots, D$. Using a model

$$P_{id} = \text{logit}(p_{id}) = \mu_d + x_i' b, \tag{2}$$

and Schwarz's criterion we identified two county-level covariates: percent of workforce in white collar jobs ($x_{i1}$) and percent of persons aged 25+ with no more than a ninth-grade education ($x_{i2}$). A second model-fitting step was used to determine whether interactions of the two county variables with the $D$ demographic groups were present. Using a logistic regression model like (2) with additional interactions of county variables and demographic groups, we again used Schwarz's criterion and by forward stepwise selection identified the best set of interactions. We found interactions of both covariates with the demographic group of 30–39 year old whites. The estimated interaction term amounted to dropping all covariate effects on this particular demographic group. As a last check, plots of average residuals were used to determine that a linear relationship with the logits appeared reasonable.

Although the county covariates used in the model selection are close to the variables that were used to design sampling strata, Malec et al. [10] found that strata effects may still be present in the data. Sampling strata were defined by grouping the county-based PSUs into 198 strata so that the PSUs in each stratum have similar summary measures of socio-economic status [11]. By assuring that each stratum contains a prespecified sample size, a more systematic coverage of the population is possible. We will include stratum effects in the model to account for this important prior knowledge based on the design. We shall use an additional subindex $_s$ in $p_{sid}$ and $x_{sid}$ to indicate strata.

In the prior specification, instead of assuming that all stratum effects are distributed i.i.d., we allow their variability to be distinct within broad mega strata.

We defined two mega-strata by grouping the 198 strata into those representing the very large metropolitan areas (like the New York City area) versus all others.

## 4. A semi-parametric randomized block model

Based on the discussion above, we include random stratum effects $\nu_s$ as well as random (county by demographic) group effects $\beta_{sid}$. Let $n_{sid}$ denote the number of individuals interviewed in demographic domain $d$, $d = 1, \ldots, D$, county $i$, $i = 1, \ldots, I$, and stratum $s$, $s = 1, \ldots, S$. Counties are nested within strata, and demographic domains are crossed with counties.

Let $y_{sid} \sim Bin(n_{sid}, p_{sid})$ denote the number of positive responses among these $n_{sid}$ individuals. Let $s(i)$ denote the stratum containing county $i$, and let $\beta_i = (\beta_{sid},\ s = s(i),\ d = 1, \ldots, D)$ denote the $D$ dimensional random effects vector for county $i$. Since county indices $i$ are unique across stata, the subindex $_s$ in $\beta_{sid}$ is redundant and we use it only when it helps to clarify the hierarchical structure of the model. We assume a logistic regression of the success probabilities $p_{sid}$ on some quantitative covariates $x_{sid}$ and county-specific and stratum-specific random effects:

$$P_{sid} = \text{logit}(p_{sid}) = x'_{sid}b + \beta_{sid} + v_{s(i)}, \tag{3}$$

with priors

$$\beta_i \sim h(\beta_i),\quad v_s \sim N(0, \delta^2_{m(s)}),\quad b \sim N(m_b, V_b),\quad \delta_m^{-2} \sim \text{Ga}(a_\delta, b_\delta)$$

$m = 1, \ldots, M$. Here $m(s)$ denotes the mega-stratum containing stratum $s$, and $\text{Ga}(a, b)$ denotes a gamma distribution with mean $a/b$. In words, the random effects $\beta_i$ are generated by some distribution $h(\cdot)$, details of which will be discussed below. The stratum random effects are *a priori* normal with random variance $\delta_m$. The model is completed with conjugate hyperpriors for $\delta_m$ and $b$, with the latter possibly non-informative by choosing $V_b^{-1} = 0$. While a large number of experimental units are available to inform inference about the county-specific random effects $\beta_i$, only a moderate (198) and small (2) number of stratum and mega-stratum-specific effects are included in the model. Therefore, we use fully parametric random effects distributions for $\nu_s$ and $\delta_m$, but a flexible non-parametric model for $\beta_i$.

The choice of the prior model $h(\cdot)$ is guided by the following considerations. First, health related outcomes always vary by age and race, but are correlated within counties. Thus we need a multivariate prior on $(\beta_{i1}, \ldots, \beta_{iD})$ allowing for different effects in different demographic domains, and interactions between these effects. Second, the model includes only few covariates, leaving significant heterogeneity due to other un-recorded covariates. To account for such heterogeneity the model needs to allow for possible clusters of sub-populations not identified by the given covariates and overdispersion. Third, as with any health outcome the model needs to accommodate outliers without unduly influencing inference. Finally, the model should be a natural generalization of more conventional multivariate normal random effects distributions.

These considerations lead us to use a mixture of normals prior model. Let $\varphi_\theta(\cdot)$ denote a multivariate normal probability density function with moments $\theta = (\mu, \Sigma)$. Then

$$\beta_i \sim \sum_{j=1}^{\infty} w_j N(\beta_i; \underbrace{\mu'_j, \Sigma'_j}_{\theta'_j}) = \int \varphi_\theta(\beta_i) dG(\theta),$$

where

$$G = \sum_{j=1}^{\infty} w_j \delta_{\theta'_j}.$$

The mixture of normals model (4) allows for heterogeneity, outliers, skewness etc., as desired. The model includes a simple multivariate normal prior model as a special case. By choosing a hyperprior on $G$ which *a priori* favors a few dominating terms in the mixture we formalize the idea that *a priori* we assume simple structure, but as the data dictates the model allows introduction of more complicated structure *a posteriori*, like a discrete mixture of a few dominating normal kernels. This is achieved using a DP prior on $G$, $G \sim DP(\alpha G_\nu)$. Here $G_\nu$ is the (standardized) base measure and $\alpha > 0$ is the total mass parameter. The base measure can possibly depend on further hyperparameters $\nu$. Below, we will specify a base measure and hyperprior on $\nu$ and the total mass parameter $\alpha$. See Ferguson [4] for a complete description of the DP; a discussion of DP and DP mixture models like (4) can be found in, e.g., [6], [13] and [23].

We summarize some properties which are relevant in the context of our application.

1. The base measure $G_\nu$ has an interpretation as the prior mean for the random measure; the total mass parameter $\alpha$ is a precision parameter. For any measurable set $A$ we have

$$E\{G(A)\} = G_\nu(A) \text{ and } Var\{G(A)\} = G_\nu(A)\{1 - G_\nu(A)\}/(1+\alpha).$$

2. The random measure $G$ is a.s. discrete. Let $F_0$ and $F$ denote the c.d.f. of $G_\nu$ and $G$, respectively. Then $F$ can be thought of as a random step function approximating $F_0$. The size $w_j$ of the steps depends on $\alpha$. The larger the $\alpha$, the smaller the weights, i.e., steps sizes, $w_j$.
3. Consider a random sample $\theta_i \sim G$, $i = 1, \ldots, n$. Because of the discreteness of $G$, with positive probability some of the $\theta_i$ will be identical. Specifically, we have the prior probabilities $P(\theta_i = \theta_j | \theta_h) = 1/(\alpha + n - 1)$ for $j < i$, $i \notin \{j, h\}$.

A commonly used device in posterior simulation when the likelihood involves a mixture like in (4) is to break the mixture by introducing latent variables [18]. We implement this by rewriting the DP mixture model (4) as

(4) $$\beta_i \sim N(\beta_i; \mu_i, \Sigma_i), \quad (\mu_i, \Sigma_i) \sim G, \quad G \sim DP(\alpha, G_\nu).$$

The model is completed by specifying a base measure $G_\nu$,

(5) $$G_\nu(\mu, \Sigma) = N(\mu; m, B) \, W[\Sigma^{-1}; s, (sS)^{-1}],$$

and a hyperpior for $\nu = (m, B, S)$ and $\alpha$,

(6) $$S \sim W[q, (R/q)], \quad m \sim N(a, A), \quad B^{-1} \sim W[c, (cC)^{-1}], \quad \alpha \sim G(a_\alpha, b_\alpha).$$

Here $W(n, A)$ denotes a Wishart distribution with scalar parameter $n$ and matrix parameter $A$, and $G(a, b)$ denotes a Gamma distribution with shape parameter $a$ and scale parameter $b$, parameterized such that the expected value of $G(a, b)$ is $a/b$.

The model is summarized in Figure 1.

We will refer to model (3) together with prior (4) or, equivalently, (4) and hyperprior (5) and (6) as the MDP (Dirichlet process mixture) model. Alternatively to the MDP prior one could choose fully parametric normal mixture models as proposed by, for example, Roeder and Wasserman [19] or Richardson and Green [17]. Based on experience with similar models, we would expect predictive inference, like the posterior predictive inference for state totals, including counties without samples, not to be much different under these models than inference using mixture models with DP priors. Still, we prefer the semi-parametric formulation for several reasons. First, the prior parameters $G_\nu$ and $\alpha$ in the DP prior model have a straightforward interpretation as the mean and dispersion parameters ("equivalent prior sample size"). It is not immediately clear what implication certain choices for priors on the parameters of the mixture model would have on the marginal distribution of $y$ in the finite mixture model. Second, the computational effort is comparable in all three models.

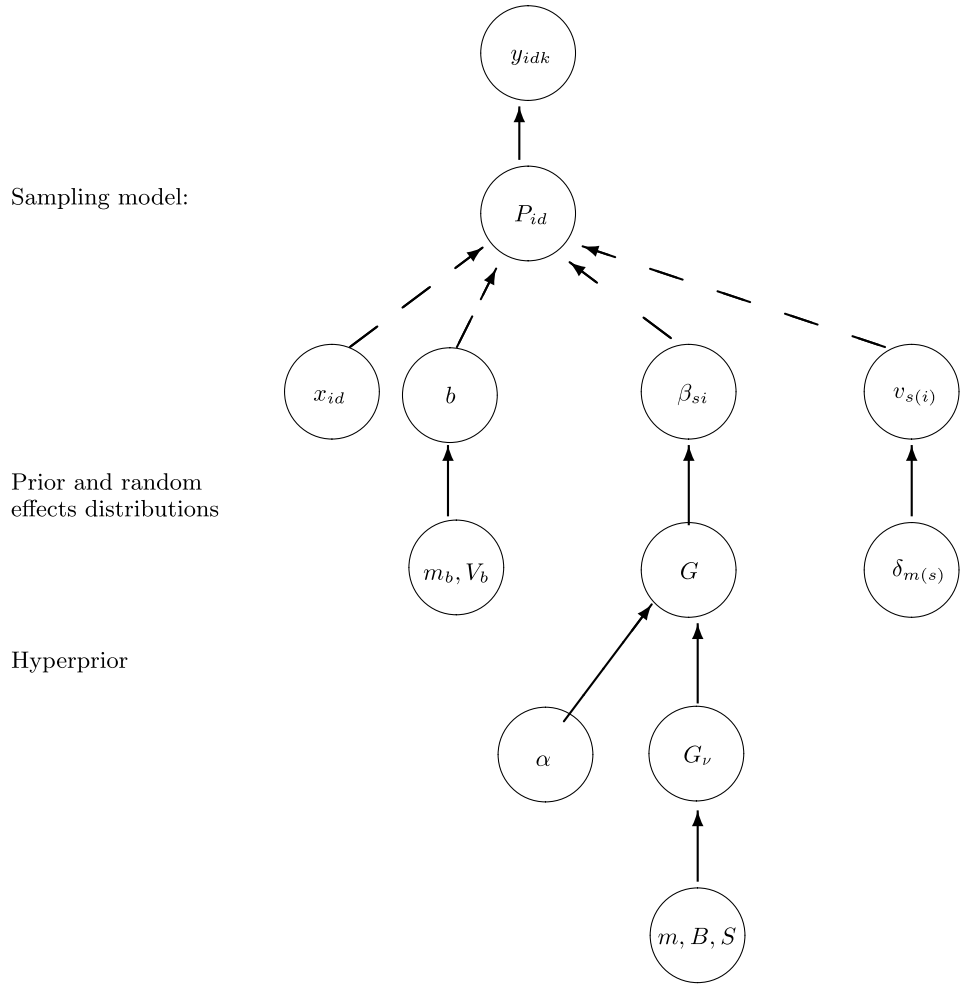

FIG 1. *Graphical model representation. Circles indicate the variables, including the data, parameters, covariates and hyperparameters. Solid arrows indicate that the probability models for the target variable are indexed by the variable at the sources of the arrow. Dashed lines represent the deterministic expression for $P_{sid}$. Fixed hyperparameters are not indicated.*

## 5. Posterior simulation

Model (3) with prior (4)–(6) can be estimated by Markov chain Monte Carlo simulation. We shall use $\beta$ for $\beta = (\beta_i, i = 1, \ldots, I)$, $\mu$ for $\mu = (\mu_1, \ldots, \mu_I)$, etc. Let $y$ denote the data vector, $y_i = (y_{sid}, d = 1, \ldots, D)$ denote the data vector for county $i$ (with $s = s(i)$ equal to the stratum containing county $i$), and $y_s = (y_{sid}, i : s(i) = s, d = 1, \ldots, D)$ denote the data vector for stratum $s$. An entry of the type $a|b, c, y$ indicates that parameter $a$ is being updated conditional on the currently imputed values of $b$ and $c$ and the data $y$. Absence of a variable $d$ in the conditioning set indicates conditional independence of $a$ and $d$ or that the model is being marginalized over $d$. We outline the sequence of the updating scheme, with details discussed below.

Let

(i) $b|\beta, v, y$,  (ii) $\beta_i|\mu_i, \Sigma_i, b, v, y_i$,  (iii) $v|b, \beta, \delta, y$,  (iv) $\delta|v$,  (v) $\mu, \Sigma, \nu, \alpha|\beta$.

Steps (i)–(v) describe the transition probability of a Markov chain in $(b, \beta, v, \mu, \Sigma)$. Step (v) refers to updating the parameters of the mixture model $h(\cdot)$, including $\mu_i, \Sigma_i, i = 1, \ldots, I$, the hyperparameters $\nu$ of the base measure $G_\nu$, and the total mass parameter $\alpha$. In a straightforward Gibbs sampler implementation each of the updating steps would draw from the respective complete conditional posterior distribution to generate a new imputed value of the respective parameter. Unfortunately this is only possible for Step (iv). In all other steps the conditional posterior distribution is not in a format allowing efficient random variate generation. The appropriate MCMC implementation for steps (i)–(iii) is explicated in the Appendix. Step (v) is described in [15].

By construction, the stationary distribution of this chain is the desired posterior distribution $p(b, \beta, v, \mu, \Sigma|y)$. Most posterior inferences take the form of integrals with respect to the posterior, such as the posterior mean $\bar{b} = \int b \, dp(b, \beta, v, \mu, \Sigma|y)$, and such posterior integrals can be approximated by ergodic averages. For example $\bar{b} \approx 1/T \sum_{t=1}^{T} b^t$, where $b^t$ denotes the value of $b$ after $t$ iterations of the Markov chain.

The aim of the small area estimation model (3) is to provide inference for both the sampled units and for the sampling units for which no data is available, and to summarize such inference by subpopulations of interest, like states, etc. The posterior simulation described earlier in this section allows us to compute such inference with minimal additional computational cost. Index with $i = I + 1, \ldots, J$ counties not included in the available sample. Denote with $N_{sid}, i = 1, \ldots, J, s = s(i), d = 1, \ldots, D$, the total populations in each cell of county and demographic domain. The population totals $N_{sid}$ and covariates $x_{sid}$ are available for all counties $i = 1, \ldots, J$, from census data. To compute inference for state totals, we proceed by the following steps. Index the states with $a = 1, \ldots, A$. Denote with $a(i)$ the index of the state $a$ containing county $i$. Let $Y_{sid}$ denote the *total* number of subjects in county $i$ and demographic domain $d$ who had a mammogram within the last two years. Let $Y_a = \sum_{\{i:a(i)=a\}} \sum_d Y_{sid}$ denote the total for state $a$.

At each iteration of the Markov chain Monte Carlo simulation we have imputed values for $b$, $\beta_{sid}, i = 1, \ldots, I$, and $v_s, s = 1, \ldots, S$. To impute random effects for counties absent in the sample, $i = I + 1, \ldots, J$, we simulate values for $\theta_i = (\mu_i, \Sigma_i)$, $i = I + 1, \ldots, J$, using the following probabilities:

$$P(\theta_i = \theta_j) = 1/(\alpha + i - 1), \quad j = 1, \ldots, i - 1,$$
(7) $\quad P(\theta_i \neq \theta_j, j < i) = \alpha/(\alpha + i - 1)$ and $P(\theta_i|\theta_i \neq \theta_j, j < i) = G_\nu(\theta_i),$

and $P(\beta_i|\theta_i) = N(\mu_i, \Sigma_i)$. Given imputed values for the random effects for *all* counties and all strata, and for the logistic regression coefficients $b$, we can now impute success probabilities $p_{sid}$ for all counties, and simulate total counts $Y_{sid} \sim Bin(N_{sid}, p_{sid})$ for all counties. Adding up the state totals $Y_a = \sum_{i:a(i)=a} \sum_d Y_{sid}$ we get simulated values $Y_a \sim p(Y_a|y)$ from the posterior distribution on the state totals.

## 6. The national health interview survey

Figure 2 shows the final inference on the state totals (as a percentage of total population $N_a$ for state $a$). Note how inference from the semi-parametric model corresponds to a compromise between the oversmoothing synthetic estimates, and the overfitting empirical means (based on observed data in each state only). The states are sorted by decreasing imputed probability to facilitate comparison. The posterior standard deviations in the state percentages are between 2 and 6 percentage points. This is the uncertainty due to estimating the state totals based on the partial sample information only. Another source of error is due to evaluating the posterior means numerically by simulation only. The corresponding uncertainties are negligible relative to the inherent posterior standard deviations. Figure 3 shows the same information in a geographical map of the U.S.

Figure 4 shows a summary of the imputed random effects $\beta_{sid}$. Clearly, assuming random effects for different domains to be i.i.d. generated from some univariate random effects distribution would be inappropriate. Posterior correlations of $\beta_{sid}$ across $d$ range from −0.5 to 0.6. Figure 5 shows some features of the estimated mixing distribution $G$. A fully parametric model would correspond to either a point mass $G$, or a conjugate multivariate normal (in $\mu$) measure $G$, corresponding to a hierarchical model. Neither seems to be a good approximation to $G$. Figure 6 shows the posterior distribution on the number of distinct $\mu_i$ in the mixture model (4). The estimated stratum-specific random effects $v_s$ are shown in Figure 7.

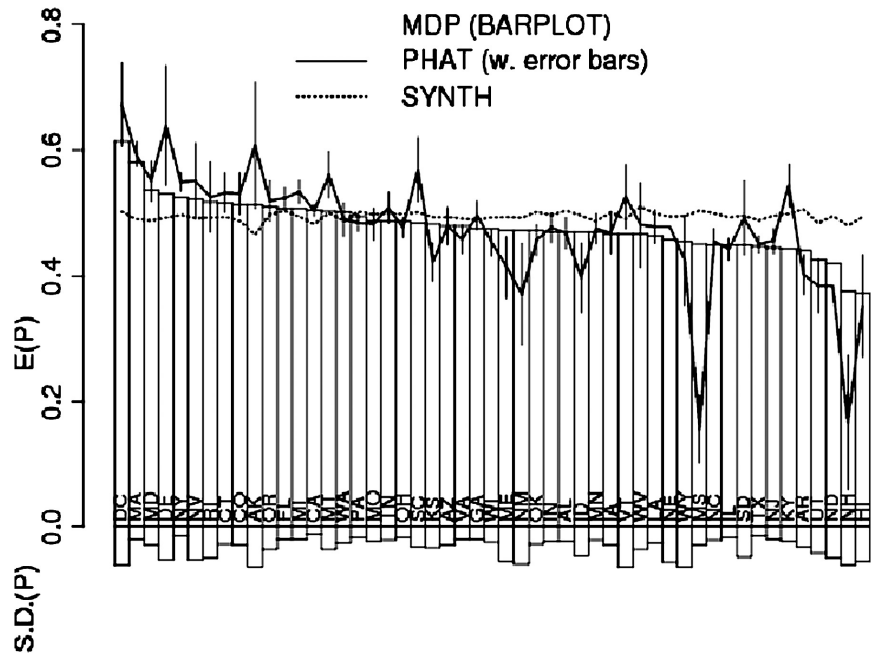

FIG 2. *Estimated state totals $E(Y_a|y)$ under the proposed model (bar plot), the synthetic method (dashed line) and as sample averages over observed samples in each state (dotted line). The short bars below the horizontal axis show one posterior s.d. $SD(Y_a|y)$ for the state totals. Error bars at each sample average estimate indicate corresponding sampling errors (assuming that all observations in a given state were independent). The sample did not include any data for NE and ND. Thus, there is no "sample estimate" for these states.*

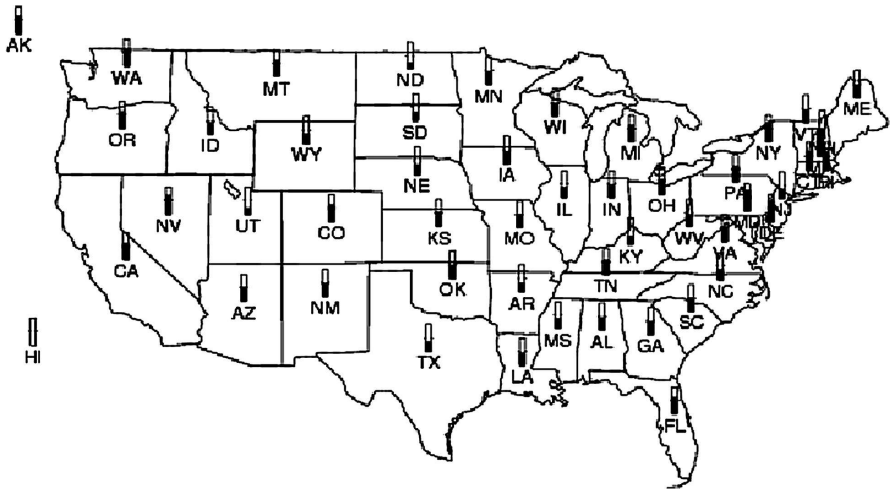

Fig 3. *Posterior predictive means per state for the percentage of women who had a mammogram done in the past two years. The solidly filled fraction of the thermometer in each state corresponds to the estimated mammogram utilization in that state. To highlight the differences between states, the measurements are shown relative to the minimum and maximum estimated percentage usage. A fully filled thermometer corresponds to 58.4%, an entirely empty thermometer corresponds to 37%.*

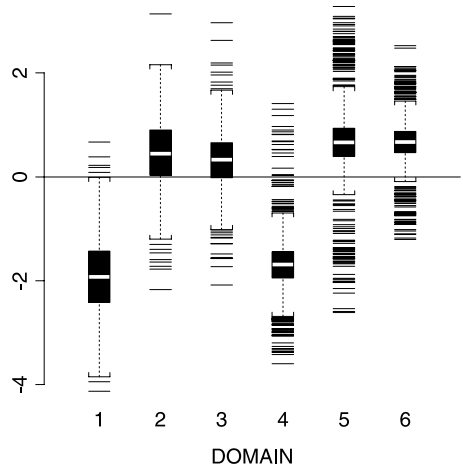

Fig 4. *Boxplots of imputed random effects $\beta_{sid}$, split by domain d, based on the random effects as imputed after 3000 iterations of the simulation.*

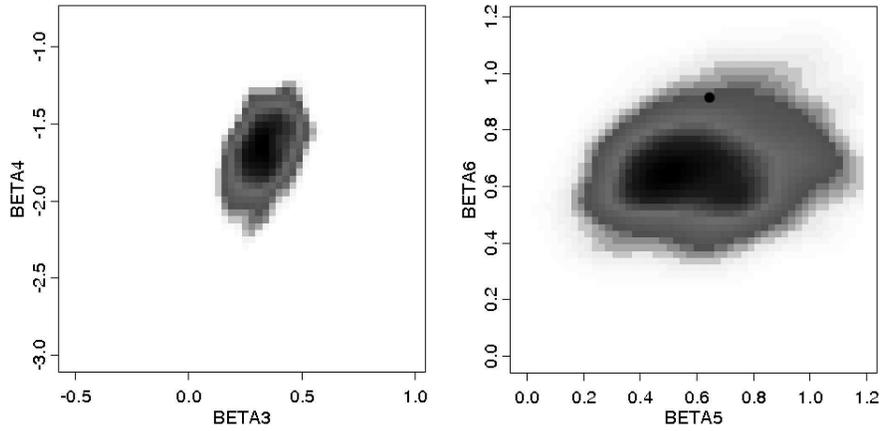

Fig 5. *Estimated mixing distribution $G(\mu)$.*

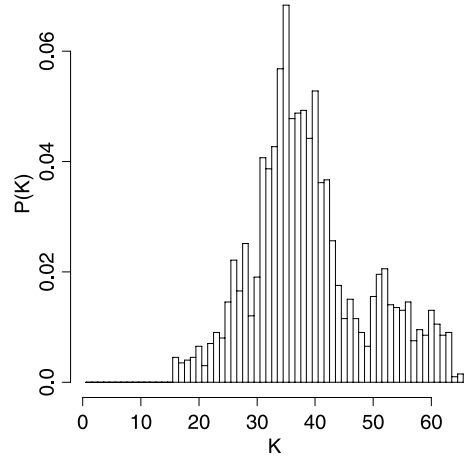

Fig 6. *The posterior distribution $p(k|y)$ on the number $k$ of distinct $\mu_i$ in the mixture model* (4). *There is a total of $I = 3060$ counties in the sample, i.e., $1 \leq k \leq 3060$.*

## 7. Discussion

The proposed approach addressed some limitations of currently used models in small area estimation by substituting a non-parametric model for the random effects distribution of the county- and domain-specific random effects. Resulting inference allows more fidelity to the data than oversmoothing synthetic estimates, without completely abandoning the borrowing of strength across counties, domains and strata as formalized in the hierarchical model. Still, several assumptions remain in the model. This includes the linear regression implied by the term $x'_{sid}b$ in (3) and the normal prior on the stratum-specific random effects. While residual plots (not shown here) indicate that the linear regression assumption was not severely contraindicated by the data, a less restrictive approach is desirable. Given the massive data available to fit the model, more general non-parametric regression models are possible. Also, the skewed distribution in Figure 7 indicates the need for less restrictive prior models on the stratum-specific random effects $v_s$. Both generalizations are possible with methods discussed in this paper and will be pursued in future research.

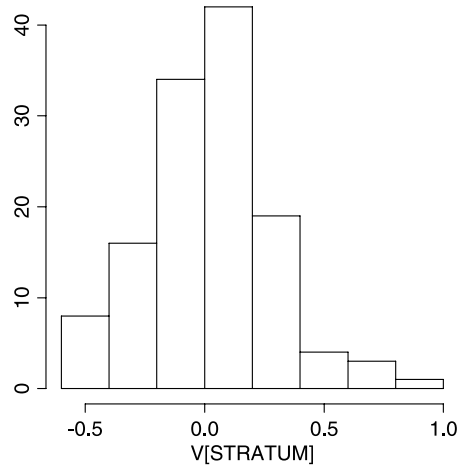

Fig 7. $E\{v_s|y\}$, $s = 1,\ldots,S$. The figure shows a histogram of imputed stratum-specific effects $v_s$.

Other extensions could add additional robustness by, for example, replacing the informal variable selection that we used for the demographic covariates by model averaging, by allowing stratum-specific random effects for the regression on demographic covariates, and by including spatial smoothing. Besides the basic test of fit by residual plots one could carry out specific model validation to investigate the use of such model extensions.

## Appendix: Resampling the logistic regression parameters

Steps (i), (ii) and (iii) in Section 5 require the updating of logistic regression parameters. We describe an independence chain implementation for Step (i). Steps (ii) and (iii) allow analogous implementations.

To update $b = (b_1,\ldots,b_p)$, we first compute the moments of the normal approximation to the likelihood based on the linearized logistic link function in a logistic regression model (cf., [12], p. 41). Denote the moments of the resulting $p$-dimensional normal approximation by $(m_1, V_1)$.

The $N(m_1, V_1)$ approximation of the likelihood is now combined with the $N(m_b, V_b)$ prior to obtain a bivariate normal approximation of the conditional posterior $p(b|\beta, v, D)$. Let $m_3$ and $V_3$ denote the moments of this approximation. Draw a candidate $\tilde{b} \sim N(m_3, V_3)$, and compute the following acceptance probability. Let $\omega$ denote the full parameter vector, and $\tilde{\omega}$ the parameter vector with $b$ replaced by $\tilde{b}$.

$$(8) \qquad a(b, \tilde{b}) = \min\left\{1, \frac{p(y|\tilde{\omega})\ \varphi(b; m_1, V_1)}{p(y|\omega)\ \varphi(\tilde{b}; m_1, V_1)}\right\},$$

where $\varphi(x; m, V)$ denotes a normal density with moments $m$ and $V$, evaluated at $x$. Replace $b$ by $\tilde{b}$ with probability $a$.

To derive expression (8) consider the general expression for acceptance probabilities in a Metropolis–Hastings algorithm (see, for example, [21]). Denote with $g(\tilde{\omega}|\omega)$ the proposal distribution. Then

$$a(\omega, \tilde{\omega}) = \min\left\{1, \frac{p(\tilde{\omega}|y)}{p(\omega|y)} \frac{g(\omega|\tilde{\omega})}{g(\tilde{\omega}|\omega)}\right\}.$$

Substitute for $g(\tilde{\omega}|\omega)/g(\omega|\tilde{\omega})$:

$$\frac{\varphi(\tilde{\beta}_i; m_3, V_3)}{\varphi(\beta_i; m_3, V_3)} = \frac{\varphi(\tilde{\beta}_i; m_1, V_1)\varphi(\tilde{\beta}_i; m_2, V_2)}{\varphi(\beta_i; m_1, V_1)\varphi(\beta_i; m_2, V_2)}$$

Also, $p(\tilde{\omega}|y)/p(\omega|y) = [p(y_i|\tilde{\omega}_i)\varphi(\tilde{\beta}_i; m_2, V_2)]/[p(y_i|\omega_i)\varphi(\beta_i; m_2, V_2)]$. The factors $\varphi(\ \cdot\ ; m_2, V_2)$ in $p(\tilde{\omega}|y)/p(\omega|y)$ cancel against the same factors in $g(\tilde{\omega}|\omega)/g(\omega|\tilde{\omega})$ and we arrive at (8).